\documentclass[twocolumn,showpacs,aps,prl,superscriptaddress]{revtex4}
\usepackage{multirow}
\usepackage{graphicx}
\usepackage{dcolumn}
\usepackage{amsmath}
\usepackage{epsfig}
\input{mybabarsym}

\newcommand{\BABARPubYear}    {09}
\newcommand{\BABARPubNumber}  {014}

\newcommand{\SLACPubNumber} {13675}
\newcommand{\LANLNumber} {0907.4566}

\def\figurebox#1#2#3{%
    \def\arg{#3}%
    \ifx\arg\empty
    {\hfill\vbox{\hsize#2\hrule\hbox to #2{\vrule\hfill\vbox to #1{\hsize#2\vfill}\vrule}\hrule}\hfill}%
    \else
    {\hfill\epsfbox{#3}\hfill}%
    \fi}

\begin{document}

\preprint{\babar-PUB-\BABARPubYear/\BABARPubNumber} 
\preprint{SLAC-PUB-\SLACPubNumber} 

\begin{flushleft}
\babar-PUB-\BABARPubYear/\BABARPubNumber\\
SLAC-PUB-\SLACPubNumber\\
arXiv:\LANLNumber\ [hep-ex]\\[10mm]
\end{flushleft}
\title{
{\large \bf
\boldmath
Observation of the baryonic \B-decay $\Bzb \ra \LCp \antiproton \Km \pip$
\unboldmath
}
}

%% author list as of 03-Apr-2009 (488 authors)
%
\author{B.~Aubert}
\author{Y.~Karyotakis}
\author{J.~P.~Lees}
\author{V.~Poireau}
\author{E.~Prencipe}
\author{X.~Prudent}
\author{V.~Tisserand}
\affiliation{Laboratoire d'Annecy-le-Vieux de Physique des Particules (LAPP), Universit\'e de Savoie, CNRS/IN2P3,  F-74941 Annecy-Le-Vieux, France}
\author{J.~Garra~Tico}
\author{E.~Grauges}
\affiliation{Universitat de Barcelona, Facultat de Fisica, Departament ECM, E-08028 Barcelona, Spain }
\author{M.~Martinelli$^{ab}$}
\author{A.~Palano$^{ab}$ }
\author{M.~Pappagallo$^{ab}$ }
\affiliation{INFN Sezione di Bari$^{a}$; Dipartimento di Fisica, Universit\`a di Bari$^{b}$, I-70126 Bari, Italy }
% ...
\collaboration{The \babar\ Collaboration}
\noaffiliation

\date{\today}

\begin{abstract}
We report the observation of the baryonic \B-decay $\Bzb \ra \LCp \antiproton \Km \pip$, excluding contributions from the decay $\Bzb \ra \LCp \Lbar \Km$.
Using a data sample of $467 \time 10^6$ \BBb pairs collected with the \babar detector at the \pep2 storage ring at SLAC, the measured branching fraction is $(4.33 \pm 0.82_{\rm stat} \pm 0.33_{\rm syst} \pm 1.13_{\LCp}) \times 10^{-5}$. 
In addition we find evidence for the resonant decay $\Bzb \ra \Sigma_c(2455)^{++} \antiproton \Km$ and determine its branching fraction to be $(1.11 \pm 0.30_{\rm stat} \pm 0.09_{\rm syst} \pm 0.29_{\LCp}) \times 10^{-5}$. The errors are statistical, systematic, and due to the uncertainty in the $\LCp$ branching fraction. For the resonant decay $\Bzb \ra \LCp \antiproton \Kstarzb$ we obtain an upper limit of $2.42 \times 10^{-5}$ at $90\%$ confidence level.
\end{abstract}


\maketitle

While $(6.8 \pm 0.6)\%$ \cite{ref:PDG} of all \B-meson decays have baryons in their final state, very little is known about the decay mechanisms behind these decays and more generally about hadron fragmentation into baryons. One way to enhance our understanding of baryon production in \B decays may be to compare decay rates to related exclusive final states.

In this paper we present a measurement of the Cabibbo-suppressed decay $\Bzb \ra \LCp \antiproton \Km \pip$ \cite{footnote}. This decay can be compared with the Cabibbo-favored decay $\Bzb \ra \LCp \antiproton \pim \pip$, which has been observed by the CLEO \cite{ref:CLEO} and Belle \cite{ref:Belle2} collaborations. The average of the branching fraction results from these two experiments are $(12.6 \pm 1.3 \pm 3.3) \times 10^{-4}$ for $\Bzb \ra \LCp \antiproton \pim \pip$ and $(2.3 \pm 0.3 \pm 0.6) \times 10^{-4}$ for the resonant subchannel $\Bzb \ra \Sigma_c(2455)^{++} \antiproton \pim$, where the first uncertainty is the combined statistical and systematic error and the second  one is the error on the $\LCp \ra \proton \Km \pip$ branching fraction. 
If only Cabibbo suppression is taken into account one expects the ratio of the corresponding Cabibbo-favored and suppressed decays to be close to $|V_{\rm us}/V_{\rm ud}|^2$, where $V_{\rm us}$ and $V_{\rm ud}$ are Cabibbo-Kobayashi-Maskawa matrix elements. Any deviation from this value indicates a contribution from the additional decay amplitudes possible in the Cabibbo-favored decays.

This analysis is based on a dataset of about $426 \invfb$, corresponding to $467 \times 10^6$ \BBb pairs, collected with the \babar detector at the \pep2 asymmetric-energy \epem storage ring, which was operated at a center-of-mass energy equal to the \FourS mass (on resonance). In addition, a dataset of $44\invfb$ collected approximately $40\mev$ below the \FourS mass (off resonance) is used to study continuum background. The \babar detector is described in detail elsewhere \cite{ref:NIM}. For simulated events we use EvtGen \cite{ref:EvtGen} for the event generation and GEANT4 \cite{ref:geant} for the detector simulation.

For the decay $\LCp \ra \proton \Km \pip$ a vertex fit is performed and the invariant mass is required to fall in the interval $2.277 < m_{\proton \kaon \pion} < 2.295 \gevcc$.
For the reconstruction of the \B-candidate, the mass of the \LCp-candidate is constrained to the nominal mass of the \LCp \cite{ref:PDG} and is combined with \antiproton, \Km, and \pip candidates. Afterwards the whole decay tree is fitted to a common vertex and the $\chi^2$ probability of this fit is required to exceed $0.2\%$.

The selection of proton, kaon, and pion candidates is based on measurements of the specific ionization in the silicon vertex tracker and the drift chamber, and of the Cherenkov radiation in the detector of internally reflected Cherenkov light \cite{Aubert:2002rg}.
The proton and antiproton selection uses in addition information from the electromagnetic calorimeter. The average efficiency for pion identification is about $95\%$ while the typical misidentification rate is $10\%$, depending on the momentum of the particle. The efficiency for kaon identification varies between $60\%$ and $90\%$ while the misidentification rate is smaller than $5\%$. The efficiency for proton and antiproton identification is about $90\%$ with a misidentification rate around $2\%$.   

The separation of signal and background of the candidate sample is obtained using two kinematic variables, $\DeltaE = E_{\B}^* - \sqrt{s}/2$ and $\mes = \sqrt{(s/2+\mathbf{p}_i \cdot \mathbf{p}_{\B})^2/E_i^2-\left|\mathbf{p}\right|_{\B}^2}$. Here, $\sqrt{s}$ is the initial center-of-mass energy, $E_{\B}^*$ the energy of the \B-candidate in the center-of-mass system, $(E_i, \mathbf{p}_i)$ is the four-momentum vector of the \epem system and $\mathbf{p}_{\B}$ the \B-candidate momentum vector, both measured in the laboratory frame.
For true \B-decays \mes is centered at the \B-meson mass and \DeltaE is centered at zero. Throughout this analysis, \B-candidates are required to have an \mes value between $5.275$ and $5.286 \gevcc$.

After applying all selection criteria there are on average $1.16$ candidates per event. If the \B-candidates have different \LCp-candidates we select the one with the invariant $\proton \Km \pip$ mass closest to the nominal \LCp mass \cite{ref:PDG}. If the candidates share the same \LCp we retain the one with the best vertex fit.

\begin{figure}
	\centering\includegraphics[width=.5\textwidth]{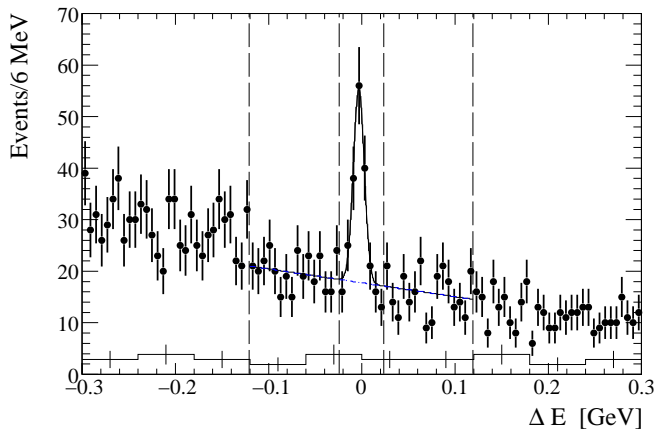}
	\caption{Fitted \DeltaE distribution in data with all selection criteria applied (data points). Shown are all reconstructed events $\Bzb \ra \LCp \antiproton \Km \pip$. The continuum background, described by off-resonance data, is overlaid (histogram). The dashed, vertical lines indicate the signal- and sideband.}
	\label{fig:DeltaE_all}
\end{figure}

\begin{figure}
	\centering\includegraphics[width=.5\textwidth]{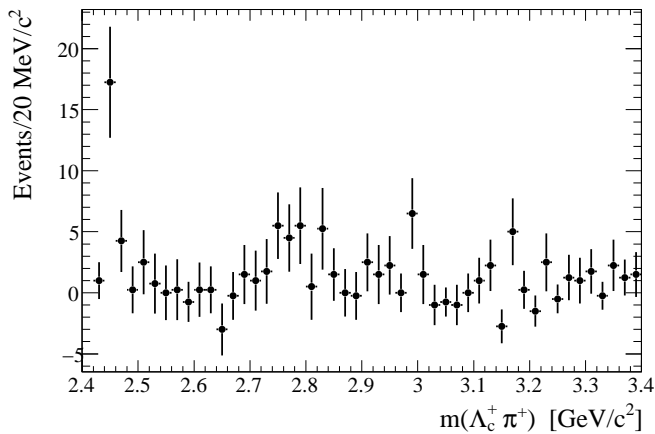}
	\caption{Invariant $\LCp \pip$ mass in data with the \DeltaE sideband subtracted. A clear $\Sigma_c(2455)^{++}$ signal is visible.}
	\label{fig:SigmaC}
\end{figure}

\begin{figure}
	\centering\includegraphics[width=.5\textwidth]{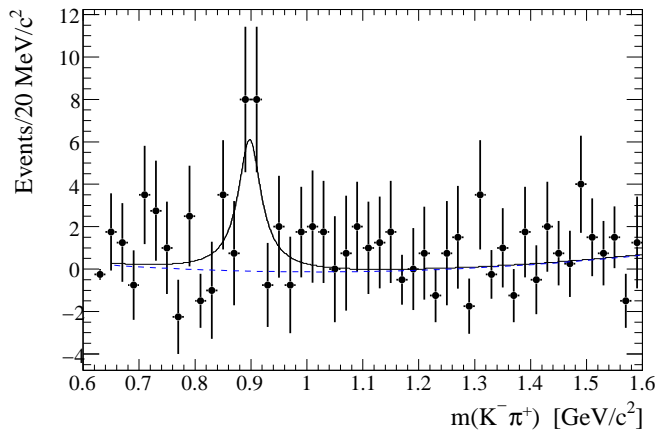}
	\caption{Sideband subtracted invariant $\Km \pip$ mass with the $\Sigma_c(2455)^{++}$ signal region ($2.447<m(\LCp \pip)<2.461\gevcc$) excluded. The solid curve is the fit, which is the sum of a nonrelativistic Breit-Wigner function and a second order polynomial. The dashed curve is the parabolic portion. An enhancement at the \Kstarzb mass of $896 \mevcc$ is visible.}
	\label{fig:Kstar}
\end{figure}

The significance of the $\Bzb \ra \LCp \antiproton \Km \pip$ signal is determined from a fit to the observed \DeltaE distribution (see Fig. \ref{fig:DeltaE_all}). As the fit function we use a straight line for background and a Gaussian for signal. Fitting between $-0.12 \gev$ and $0.12 \gev$ we obtain $82 \pm 17$ signal events and determine a significance of $8.8$ standard deviations for this decay.
Here, and in the following, we calculate the significance as the square root of the difference of 2 times the log likelihood of a fit with and without signal component. 
Like the Cabibbo-favored decay $\Bzb \ra \LCp \antiproton \pim\pip$ the Cabibbo-suppressed decay can proceed via different resonant subchannels.
Figures \ref{fig:SigmaC} and \ref{fig:Kstar} show the sideband subtracted $\LCp \pip$ and $\Km \pip$ invariant mass distributions, respectively. 
Here, the signal region corresponds to $|\DeltaE|<0.024\gev$, and the sideband regions to $0.024<|\DeltaE|<0.12\gev$.
We find evidence for the decay $\Bzb \ra \Sigma_c(2455)^{++} \antiproton \Km$ ($4.3\sigma$) and hints on the decay $\Bzb \ra \LCp \antiproton \Kstarzb$ ($2.7\sigma$). For the determination of the significance we use in both cases a second order polynomial for background and as signal function we use in the first case a Gaussian and in the latter a nonrelativistic Breit-Wigner.

For the determination of the efficiency corrected signal yield, we divide the phase space into smaller regions. In order to account for the resonant substructure, the following regions are used:
\renewcommand{\labelenumi}{\arabic{enumi})\;}
\begin{enumerate}
	\item The $\Sigma_c(2455)^{++}$ signal region in the range from $2.447$ to $2.461 \gevcc$ in $m(\LCp \pip)$,
	\item the \Kstarzb signal region from $0.8$ to $1.1 \gevcc$ in $m(\Km \pip)$, excluding region $1)$, and
	\item all events that are not in region $1)$ or $2)$.
\end{enumerate}
The events in region 3 show no further significant resonant structure, but are also not uniformly distributed in phase space. Since we use a phase space model in our Monte Carlo simulation we correct the efficiency as a function of $m(\LCp \antiproton \pip)$.
We determine the signal yield in the different regions by subtracting the extrapolated background from the observed number of \B-candidates in the \DeltaE signal region. The background is determined with a linear fit to the \DeltaE distribution in the \DeltaE sidebands, $0.024<|\DeltaE|<0.12\gev$.
For the efficiency estimation we use the same fit strategy as for the signal yields, but instead of a straight line we use a second order polynomial as fit function to account for the small combinatoric background in the signal Monte Carlo simulation. Here, we use nonresonant Monte Carlo events for regions 2 and 3 and for region 1 we use $\Bzb \ra \Sigma_c(2455)^{++} \antiproton \Km$ Monte Carlo events since this region is almost saturated by resonant events.
The number of signal events $N_{\rm sig}$, as well as the efficiencies $\varepsilon$, for the three regions are listed in Table~\ref{tab:eff}.
\begin{table}
	\caption{Number of signal events, $N_{\rm sig}$, and efficiencies $\varepsilon$ for the three regions used to obtain the signal yield.}
	\begin{tabular}{lcc}\hline
		Region			& $N_{\rm sig}$		& $\varepsilon$		\\\hline
		1 ($\Sigma_c^{++}$)	& $17.3 \pm 4.6$	& $(6.64 \pm 0.04)\%$	\\
		2 (\Kstarzb)		& $26.5 \pm 9.7$	& $(8.60 \pm 0.07)\%$ 	\\
		3			& $39.7 \pm 12.2$	& $(8.94 \pm 0.25)\%$	\\
		\hline
	\end{tabular}
	\label{tab:eff}
\end{table}
Using these values the overall branching fraction is calculated as
\begin{align}
	\nonumber &\BR(\Bzb \ra \LCp \antiproton \Km \pip) \\
	\nonumber &\quad = \frac{1}{\BR(\LCp \ra \proton \Km \pip) \cdot N_{\BBb}} \cdot \sum_{i=1}^{3} \frac{N_{{\rm sig},\,i}}{\varepsilon_i} \\
	&\quad= (4.33 \pm 0.82_{\rm stat} \pm 1.13_{\LCp})\times 10^{-5} \label{eq:bf_overall}
\end{align}
with $\BR(\LCp \ra \proton \Km \pip) = (5.0 \pm 1.3) \%$ \cite{ref:PDG} and $N_{\BBb} = N_{\Bzb} + N_{\Bz} = (467 \pm 5)\times 10^6$, assuming equal production of \BzBzb and \BpBm in the decay of the \FourS. In Eq. (\ref{eq:bf_overall}) and in the following branching fractions, the first uncertainty is statistical, while the second one arises from the branching fraction of the \LCp. 
The final state $\LCp \antiproton \Km \pip$ may also include contributions from the decay $\Bzb \ra \LCp \Lbar \Km$. Our cut on the vertex fit probability, however, would strongly suppress this contribution, hence the branching fraction (\ref{eq:bf_overall}) is understood to not include this decay. This is corroborated by the fact that the $\antiproton\pip$ invariant mass distribution shows no $\Lbar$ peak.

\begin{table}
	\caption{Number of signal events, $N_{\rm sig}$, and the efficiency $\varepsilon$ for the resonant decays via the $\Sigma_c(2455)^{++}$ and the \Kstarzb.} 
	\begin{tabular}{lcc}\hline
		Resonance		& $N_{\rm sig}$		& $\varepsilon$		\\\hline
		$\Sigma_c(2455)^{++}$	& $16.0 \pm 4.3$	& $(6.15 \pm 0.04)\%$	\\
		\Kstarzb		& $20.9 \pm 7.9$	& $(8.38 \pm 0.05)\%$	\\\hline
	\end{tabular}
	\label{tab:eff_res}
\end{table}

For the $\Sigma_c(2455)^{++}$ subchannel we determine the signal yield with a fit to the \DeltaE sideband subtracted $m(\LCp \pip)$ distribution. 
We obtain the signal yield by subtracting from the number of events observed in the $\Sigma_c(2455)^{++}$ signal region the background yield extrapolated from a fit of a second order polynomial to the $\Sigma_c(2455)^{++}$ mass sidebands. 
Here, the signal region is defined as $2.447<m(\LCp \pip)<2.461\gevcc$ while the mass sidebands are $2.426<m(\LCp \pip)<2.447\gevcc$ and $2.461<m(\LCp \pip)<2.7\gevcc$.
The efficiency is estimated by using the same fit strategy on $\Bzb \ra \Sigma_c(2455)^{++} \antiproton \Km$ Monte Carlo events. Both, the signal yield as well as the efficiency for this resonant subchannel are given in Table \ref{tab:eff_res}. Using these values we obtain a branching fraction of $(1.11 \pm 0.30_{\rm stat} \pm 0.29_{\LCp})\times 10^{-5}$ for this subchannel, under the assumption that the $\Sigma_c(2455)^{++}$ decays entirely into $\LCp \pip$.

For the \Kstarzb subchannel we determine the signal yield by a fit to the \DeltaE sideband subtracted $m(\Km \pip)$ distribution, excluding the $\Sigma_c(2455)^{++}$ signal region. 
Here, we use the sum of a second order polynomial and a nonrelativistic Breit-Wigner function in the range from $0.64$ to $1.6\gevcc$ as the fit function. The nonrelativistic Breit-Wigner distribution is added in order to get a proper background description from the fit. For the fit we fix the width and the mean of the Breit-Wigner function to its measured values \cite{ref:PDG}, and determine the signal yield by subtracting the integral of the background function between $0.8$ and $1.0\gevcc$ from the number of events in this region.
The efficiency is estimated applying the same fit procedure to $\Bzb \ra \LCp \antiproton \Kstarzb$ Monte Carlo events. With the obtained values, which are listed in Table \ref{tab:eff_res}, we estimate a branching fraction of $(1.60 \pm 0.61_{\rm stat} \pm 0.42_{\LCp}) \times 10^{-5}$ for this subchannel taking into account that $2/3$ of the \Kstarzb decay into $\Km\pip$. 

Several sources of systematic uncertainties have been investigated. Most of these are derived from studies of data control samples and by comparison between data and Monte Carlo events. The systematic uncertainties arise from the reconstruction of charged tracks ($1.4\%$), the charged particle identification ($2.4\%$), and the number of \BBb pairs ($1.1\%$).
The uncertainty due to the \DeltaE background parametrization in data is determined by extracting the signal yield with a second order polynomial instead of a straight line ($4.7\%$). The influence of the signal- and sideband definitions is estimated by changing their definitions to $|\DeltaE|<0.036\gev$ and $0.036<|\DeltaE|<0.12\gev$, respectively, and extracting the signal yields with these new definitions ($3.3\%$). A further systematic uncertainty is the phase space model used for the Monte Carlo simulation ($1.0\%$), which is determined by reweighting the Monte Carlo events to match the observed $m(\LCp \antiproton \pip)$ distribution in data. In order to estimate the uncertainties arising from the applied $m(\LCp)$ ($3.4\%$) and $\chi^2$ probability ($0.8\%$) selection criteria we vary the criteria by $0.5\mevcc$ and $0.001$, respectively. The overall systematic uncertainty is $7.5\%$.

For the low significance $\LCp \antiproton \Kstarzb$ signal, we determine an upper limit of $2.42 \times 10^{-5}$ at $90\%$ confidence level. This limit is calculated assuming a Gaussian a posteriori probability density with $\sigma = 0.63\times 10^{-5}$ which includes statistical and systematic errors, and evaluating $90\%$ of the integral in the physical region.

In summary, we observe the decay $\Bzb \ra \LCp \antiproton \Km \pip$ with a significance of $8.8\sigma$ and measure a branching fraction of
\begin{align}
	& \nonumber \BR(\Bzb \ra \LCp \antiproton \Km \pip)\\
	& \quad = (4.33 \pm 0.82_{\rm stat} \pm 0.33_{\rm syst} \pm 1.13_{\LCp})\times 10^{-5}.
\end{align}
The ratio of the branching fraction of this decay to that of $\Bzb \ra \LCp \antiproton \pim \pip$ \cite{ref:Belle2,ref:CLEO} is $0.038 \pm 0.009$, which is smaller than $|V_{us}/V_{ud}|^2 = 0.0536 \pm 0.0020$ \cite{ref:PDG}. This is a possible indication that additional decay amplitudes for the Cabibbo-favored decay are not negligible. Here, and in the following the error on the ratio includes statistical and systematic uncertainties, while the uncertainty on the \LCp branching fraction cancels.

The branching fraction of the decay $\Bzb \ra \Sigma_c(2455)^{++} \antiproton \Km$ is determined to be
\begin{align}{}
        &\nonumber \BR(\Bzb \ra \Sigma_c(2455)^{++} \antiproton \Km)\\
        &\quad = (1.11 \pm 0.30_{\rm stat} \pm 0.09_{\rm syst} \pm 0.29_{\LCp})\times 10^{-5}.
\end{align}
The ratio of this branching fraction to that of $\Bzb \ra \Sigma_c(2455)^{++} \antiproton \pim$ \cite{ref:Belle2,ref:CLEO} is $0.048 \pm 0.016$, compatible with $|V_{us}/V_{ud}|^2$.

For the decay $\Bzb \ra \LCp \antiproton \Kstarzb$ the branching fraction is determined to be
\begin{align}{}
	&\nonumber \BR(\Bzb \ra \LCp \antiproton \Kstarzb)\\
	&\quad = (1.60 \pm 0.61_{\rm stat} \pm 0.12_{\rm syst} \pm 0.42_{\LCp}) \times 10^{-5}.
\end{align}
The $90\%$ confidence level upper limit for this decay is
\begin{align}{}
	&\BR(\Bzb \ra \LCp \antiproton \Kstarzb) < 2.42 \times 10^{-5}.
\end{align}

We are grateful for the excellent luminosity and machine conditions
provided by our \pep2\ colleagues, 
and for the substantial dedicated effort from
the computing organizations that support \babar.
The collaborating institutions wish to thank 
SLAC for its support and kind hospitality. 
This work is supported by
DOE
and NSF (USA),
NSERC (Canada),
CEA and
CNRS-IN2P3
(France),
BMBF and DFG
(Germany),
INFN (Italy),
FOM (The Netherlands),
NFR (Norway),
MES (Russia),
MEC (Spain), and
STFC (United Kingdom). 
Individuals have received support from the
Marie Curie EIF (European Union) and
the A.~P.~Sloan Foundation.


\begin{thebibliography}{99}

\bibitem{ref:PDG}
C. Amsler {\em et al.} (Particle Data Group), \pl {\bf B667}, 1 (2008).

\bibitem{footnote}
Throughout this paper, all decay   modes represent that mode and its charge conjugate.

\bibitem{ref:CLEO}
S.~A.~Dytman {\em et al.} (CLEO Collaboration),
  Phys.\ Rev.\ {\bf D66}, 091101 (2002).


\bibitem{ref:Belle2}
  K.~S.~Park {\it et al.}  (Belle Collaboration),
  Phys.\ Rev.\  D {\bf 75}, 011101 (2007).

\bibitem{ref:NIM}
B.\ Aubert {\em et al.} (\babar\ Collaboration), Nucl.\ Instrum.\ Methods {\bf A479}, 1 (2002).

\bibitem{ref:EvtGen}
  D.~J.~Lange,
  Nucl.\ Instrum.\ Meth.\  A {\bf 462}, 152 (2001).

\bibitem{ref:geant}
S. Agostinelli {\em et al.} (GEANT4 Collaboration),
Nucl. Instrum. Methods {\bf A506}, 250 (2003).

\bibitem{Aubert:2002rg}
  B.~Aubert {\it et al.}  (BABAR Collaboration),
  %``A study of time dependent CP-violating asymmetries and flavor oscillations
  %in neutral $B$ decays at the $\Upsilon(4S)$,''
  Phys.\ Rev.\  D {\bf 66}, 032003 (2002)
%%  [arXiv:hep-ex/0201020].
  %%CITATION = PHRVA,D66,032003;%%

\end{thebibliography}
\end{document}